Creation of a high spatiotemporal resolution global database of continuous mangrove forest cover for the 21st Century (CGMFC-21).

Stuart E. Hamilton* and Daniel Casey**

* Assistant Professor, Geography and Geosciences, Salisbury University, Salisbury, MD 21810. sehamilton@salisbury.edu

** Post Bachelor Fellow, Institute for Health Metrics and Evaluation, Seattle, WA. 98121. dccasey@uw.edu

**ABSTRACT**

The goal of this research is to provide high resolution local, regional, national and global estimates of annual mangrove forest area from 2000 through to 2012 with the goal of driving mangrove research questions pertaining to biodiversity, carbon stocks, climate change, functionality, food security, livelihoods, fisheries support and conservation that have been impeded until now by a lack of suitable data. To achieve this we synthesize the Global Forest Change database, the Terrestrial Ecosystems of the World database, and the Mangrove Forests of the World database to extract mangrove forest cover at high spatial and temporal resolutions. We then use the new database to monitor mangrove cover at the global, national and protected area scales. Countries showing relatively high amounts of mangrove loss include Myanmar, Malaysia, Cambodia, Indonesia and Guatemala. Indonesia remains by far the largest mangrove-holding nation, containing between 26% and 29% of the global mangrove inventory with a deforestation rate of between 0.26% and 0.66% annually. Global mangrove deforestation continues but at a much reduced rate of between 0.16% and 0.39% annually. Southeast Asia is a region of concern with mangrove deforestation rates between 3.58% and 8.08% during the analysis period, this in a region containing half of the entire global mangrove forest inventory. The global mangrove deforestation pattern from 2000 – 2012 is one of decreasing rates of deforestation, with many nations essentially stable, with the exception of the largest mangrove-holding region of Southeast Asia. We provide a standardized global spatial dataset that monitors mangrove deforestation globally at high spatiotemporal resolutions, covering 99% of all mangrove forests. These data can be used to drive the mangrove research agenda particularly as it pertains to improved monitoring of mangrove carbon stocks and the establishment of baseline local mangrove forest inventories required for payment for ecosystem service initiatives.

**INTRODUCTION**

A systematic high spatiotemporal resolution global mangrove database is lacking. Without such a database, research into mangrove functionality is on a weak empirical footing. The majority of historic mangrove cover estimates are snapshots that use aggregated data from regional or national studies. For example, the Food and Agriculture Organization of the United Nations (FAO) regularly compiles snapshots of mangrove cover at the national scale. Much of the data in these reports are singular estimates of national mangrove cover that propagate through each subsequent report and across reports. Such reports have proven important to the mangrove research community in depicting historic mangrove cover and loss but do not meet the requirements of the current mangrove research agenda that requires global mangrove data with high spatiotemporal granularity. For example, when conducting a literature search of historic mangrove cover estimates in Malaysia, Friess and Webb (2011), note that the mangrove data estimates are highly variable, and this results in high amounts of uncertainty when compiling mangrove loss trends over time. The three major issues causing this uncertainty are stated as being a lack of reporting a mangrove calculation methodology particularly in the grey literature in which mangrove cover analyses often reside; a lack of traceability of data points that comprise a study; and problematic data assumptions often due to sampling of mangroves or assumptions on the unverifiable temporal axis of a study (Friess & Webb, 2011).

Mangrove atlases provide an additional source of global mangrove cover information (e.g. Spalding *et al.*, 1997; Spalding *et al.*, 2010) and generally provide mangrove cover information at the national scale. Such atlases provide important mangrove information, particularly as it pertains to mangrove species information and the local situation of mangrove forests. Other mangrove estimates often refine either the FAO data or these atlas datasets. Using FAO and other national estimates it is noted that conflicting mangrove change trends can exist across different data sources, within the same data sources and across such significant mangrove holdings nations such as Indonesia, Brazil and the Philippines (Friess & Webb, 2014) as well as in Mexico (Ruiz-Luna *et al.*, 2008). Studies into mangrove biodiversity, mangrove functionality, mangrove carbon stocks and mangrove conservation are hindered by the conflicting trends found across these datasets (Friess & Webb, 2014). Indeed, such conflicting information hampers policy decisions not only for issues related to mangroves (Friess & Webb, 2011), but also for other forest types globally. Table 1 depicts this problematic variability within global mangrove estimates (Lanly, 1982; FAO, 1995, 1997, 2000, 2003, 2005, 2007, 2010). Depending on the datasets used, global mangrove forest cover can be represented as an increasing trend from 1980 to 2005, a decreasing trend from 1980 to 2005, or a variable trend.

Globally remotely sensed products overcome many of the caveats of national estimates from government organizations by utilizing a systematic approach to mangrove mapping   allnations. Despite this, all preexisting remotely sensed products are lacking either the spatial resolution, temporal resolution or the required mangrove classification to adequately fill the identified data gap. For example, global land cover products such as GlobCover are at 300 m resolution, are lacking a mangrove classification and have only two coverage dates post-2000. The MODIS land cover classification products are annual but also at a coarse 250 m resolution with no mangrove classification. GLC 2000 does contain a 'tree cover, regularly flooded, saline water' classification but the resolution is a coarse 1 km grid and again is a singular snapshot. The Mangrove Forests of the Word (MFW) Landsat based mangrove database overcomes many of these obstacles creating what the authors state is, "the most comprehensive, globally consistent and highest resolution (30 m) global mangrove database ever created" (Giri *et al.*, 2011 p. 154).

MFW advanced mangrove mapping by providing a systematic approach to mapping mangrove cover across all nations and thus allowing for local, regional, national and global analysis of mangrove in the year 2000. Despite this, MFW and similar global mangrove measurement models have two major limiting factors. Firstly, they lack a systematic temporal mangrove measure, as they are one-time snapshots of historic mangrove cover. Secondly, the actual measurement of mangrove at the mapping unit is presence or absence, as it does not report the actual amount of mangrove cover at each location. This may be important as mangrove forests are often fringe forests located at the terrestrial and water interface with a high likelihood that not all of the pixel area classified as mangrove may be mangrove forest. Indeed, although a mangrove stand may consistently exist over time at the pixel scale, it has been noted that the quality of the mangroves may be degraded due to pollution, grazing or oil spills, and a presence or absence approach to mangrove mapping is unlikely to capture such degradation (FAO, 2007).

Although, categorical presence and absence data is the most common form of remotely sensed forest mapping (DeFries *et al.*, 1995; Bennett, 2001), it is noted that such techniques may not represent forest heterogeneity that may be present (DeFries *et al.*, 1995) and additionally may not accurately represent true forest canopy cover (Asner *et al.*, 2005). The approach used in this manuscript likely has its highest utility when used in forest-based payment for ecosystem services (PES) programs, such as those targeted to reduce emissions from deforestation and forest degradation (REDD), which often only use forest presence or absence measures without accounting forest degradation over time. Indeed, it is noted that current remote sensing products are not adequate to capture the spatial variability required to produce accurate forest carbon maps (Asner *et al.*, 2010). In addition to systematic and annual mapping of mangrove forest, we use a percentage treecover approach to mangrove mapping as opposed to mapping based on presence or absence. That is, we report the likely amount of mangrove present at the minimum mapping unit as opposed to presence or absence of mangrove. By doing this we can capture measures of mangrove degradation and adjust for mangrove area in fringe pixel situations. The percentage cover approach is more relevant than categorical mapping methods when the mangrove analysis is concerned with measurements of standing biomass or carbon stocks as opposed to measure of biodiversity or habitat when the actual pixel cover amount may be less important.

Despite the lack of a robust post-2000 mangrove change database, concern over mangrove deforestation is well elucidated in the recent literature with numerous mangrove change studies at the global, national and local scales (e.g. Satapathy *et al.*, 2007; Hamilton, 2013). Knowledge of the economic value of mangroves to ecosystem services has existed for some time (e.g. Barbier & Cox, 2004; Barbier, 2006) with much of the literature concerned with mangrove support of fisheries (e,g, Chong, 2007; Lugendo *et al.*, 2007). Despite the important ecological services role of mangrove forest, it is in the realm of climate change that mangrove research has come to the forefront of the land-use change literature in recent years. Mangroves have been shown to contain some of the largest forest carbon sinks per hectare of any forest type globally (Bouillon *et al.*, 2008; Donato *et al.*, 2011) including substantial carbon stored below ground in mangrove soil (Donato *et al.*, 2011; Murdiyarso *et al.*, 2015). Therefore, mangrove deforestation likely releases more $CO_2$ per hectare than any other forest type. Indeed, work is underway on placing economic values of the carbon stored in mangrove forests (Siikamäki *et al.*, 2012) adding substantially to the potential economic value of preserved mangroves.

An emerging issue in the mangrove and wider forest research community is the inability of current forest databases to set baseline reference scenarios for PES schemes such as national-scale REDD projects (Angelsen *et al.*, 2012). As Table 1 indicates, utilizing FAO estimates as the baseline for REDD forest programs could result in highly unsatisfactory mangrove monitoring and evaluation. Yet, it is FAO data that is most-often used in studies concerned with the establishment of REDD baselines (e.g, Griscom *et al.*, 2009b, a; Huettner *et al.*, 2009) and compatibility with FAO is often viewed as a prerequisite of any potential REDD measure (Huettner *et al.*, 2009). The realization that the degradation portion of REDD is omitted within FAO does exist within the literature (Griscom *et al.*, 2009b). Yet, the suitability of such datasets for PES analysis appears to go mostly unaddressed, despite the realization such data has profound implications up-to-and-including the mechanisms for national participation in future climate change treaties (Angelsen, 2008).

The recently released Global Forest Cover (GFC) database (Hansen *et al.*, 2013) has the potential to overcome many of the limitations of traditional mangrove estimates stated above. It contains annual data from 2000 to 2012, as well as containing percentage tree cover at the minimum mapping unit. Unfortunately this dataset does not distinguish between forest types (Tropek *et al.*, 2014). To overcome this issue, synthesis with other datasets that define landcover at similar spatial resolutions is required.

The resolution of the mangrove data presented in this analysis is approximately 30 m, with a measures of mangrove forest cover provided at each minimum mapping unit. Our presented dataset likely contains the highest spatial resolution, highest temporal resolution and highest attribute resolution of all global mangrove datasets and allows for systematic mangrove analysis at the global, continental, country, region, estuary or even individual study area scale. Despite the importance of establishing mangrove loss trends, it is not in mangrove change analysis that these data provide the most utility but in driving research into questions related to mangrove; biomass, carbon stocks, functionality, food security, biodiversity, livelihoods and fisheries support that have been hindered until now by a lack of suitable data.

**MATERIALS AND METHODS**

To create CGMFC-21 we synthesize the GFC database (Hansen *et al.*, 2013), the MFW database (Giri *et al.*, 2011) and the Terrestrial Ecosystems of the World (TEOW) database (Olson *et al.*, 2001) in conjunction with other ancillary datasets to produce global mangrove forest cover measures for 2000 to 2012, and estimates for 2013 and 2014.

The first step in the process was to calculate year 2000 mangrove cover globally. To achieve this vector MFW was converted back into its native resolution of $2.7777 \times 10\text{-}4°$ for all locations; this resulted in a raster layer of year 2000 mangrove cover with an attribute of presence or absence. During this process, pixel alignment was enforced with GFC. Both MFW and GFC use the same native pixel size so no resampling or shifting of pixels was required. We then extracted all of the year 2000 treecover pixels that overlaid the year 2000 mangrove defined area. This resulted in only pixels that had been determined to have mangrove in the year 2000. After pixel extraction, each pixel was given an additional attribute of area in meters squared based on the percent of treecover present. This area calculation was achieved by applying a latitudinal correction to each pixel based on the Spherical Law of Cosines. This was preferable to other methods as computationally expensive reprojection was avoided and the data maintains its original coordinate system. Additionally, pixel averaging and estimating was avoided that would have resulted during data reprojection. The final step was to apply the percentage treecover value to each pixel. For example, if the pixel was determined to be 900 $m^2$ in size and it had 50% mangrove treecover then the pixel was given a mangrove value of 450 $m^2$.

Once year 2000 mangrove cover was established, GFC was queried for loss in 2001 and each loss pixel was converted into area using the methods outlined above. Pixels that had been deforested during 2001 were then integrated into the 2000 mangrove dataset to produce the 2001 dataset. This was repeated for all years from 2001 to 2012, with the preceding year becoming the baseline mangrove cover layer to establish loss for the following year. This resulted in 13 mangrove datasets (one for each year) at 2.7777 × 10-4° resolution (approximately 30 m$^2$ in the tropics) for all areas that had mangrove present in the year 2000. The 30 m global data for each year were then aggregated to the national scale with any mangrove falling outside of national boundaries being allotted to the closest nation while remaining in its actual location. The MWF mangrove measure (Supp. Table 1) is best described as monitoring mangrove forest change that has occurred in all areas that had mangrove forest present in the year 2000. It does not allow for monitoring of mangrove growth that may have occurred outside of areas that had no historic mangrove cover. This dataset is most suitable for mangrove analysis concerned with actual treecover such as aboveground and belowground biomass calculations and estimations of carbon stocks and provides a solution to the inherent issues related to establishing forest baselines for PES programs such as REDD.

The second measure of forest change focuses on forest-cover in the entire mangrove biome, as opposed to a stricter definition of verified year 2000 mangrove forests. TEOW was rasterized to 2.7777 × 10-4° resolution for all locations; this resulted in a raster layer depicting the entire mangrove biome in addition to locations with mangrove known to exist during year 2000. During this rasterization process, pixel alignment was again enforced to comply with GFC. We then extracted all of the year 2000 GFC treecover pixels that overlaid the mangrove biome pixels. This resulted in only pixels that are located within the mangrove biome or had mangrove in 2000. Again, the continuous pixel value was converted into area using a latitude adjustment grid and mangrove loss was burned into each pixel for subsequent years. As opposed to MFW, areas within the TEOW mangrove biome that had experienced mangrove forest gain were additionally added to the dataset. This mangrove measure is best described as monitoring forest change that has occurred in all areas of the mangrove biome even those outside of delineated mangrove forests. This layer allows for monitoring of mangrove growth that may have occurred outside of areas that had no historic mangrove cover. This dataset is most suitable for mangrove analysis concerned with biome characteristics such as habitat fragmentation and biodiversity analyses.

After compiling both the mangrove measures above and establishing the linearity of the mangrove change a simple OLS regression was performed on the national data to predict the mangrove areas for 2013 and 2014 and to bring the datasets to present. In addition to the global mangrove areas reported by country we extracted the data for the mangrove dominated Ramsar sites of Everglades National Park in North America, Cobourg Peninsula in Northern Australia, Sundarbans National Park on the border of India and Bangladesh, Douala Edéa National Park on the west coast of Africa in Cameroon and Cayapas-Mataje on the west coast of Ecuador bordering Colombia. We additionally calculated the mangrove deforestation trend for all protected areas globally.

To test the representativeness and accuracy of the findings presented we utilized the only other approximately 30 $m^2$ resolution measure of continuous forest cover available for one of the analysis regions. The USGS NLCD (National Land Cover Dataset) provides intermittent continuous tree cover measures for the contiguous USA (Homer *et al.*, 2012). From the 2011 NLCD data, we extracted the 2,037,420 pixels within Florida that are coincident with our 2011 mangrove data. We then converted the NLCD dataset into square meters and compared the two mangrove measures for Florida. Our dataset estimates 1341 $km^2$ of mangrove forest cover in Florida during 2011 whereas NLCD estimates 1391 $km^2$ of mangrove forest cover. The histogram of differences reflects a relatively normal distribution of difference with strong clustering around the mean difference of 25m (Supp. Methods). The 3.6% difference between the two Florida mangrove estimates increases confidence that the data presented here are an accurate and representative depiction of continuous mangrove cover that is comparable to other remote sensing derived continuous forest datasets. Additionally, a portion of the 3.6% disagreement is likely due to slightly differing sensor acquisition dates during 2011.

*Global Forest Cover*

The GFC dataset provides the most resolute global map of forest cover yet produced (Hansen *et al.*, 2013). It uses over 650,000 Landsat images to map the change in global forest cover at yearly intervals from 2000 - 2012. The dataset allows forest loss and forest gain to be measured against a baseline of year 2000 forest cover. The dataset estimates total forest loss between 2000 and 2012 of approximately 2.3 million km$^2$ with gains offsetting approximately 800,000 km$^2$ of these losses (Hansen *et al.*, 2013). Although not explicitly defined in the data, it likely captures almost all mangrove forest cover aside from juvenile mangrove forests and forests consisting wholly of mangrove scrub.

The GFC database and methodology has been criticized for not differentiating between native forest and forest plantations and ignoring the ecological role of forests. For example, it has been noted that plantation forests that displace indigenous or other more diverse forest types (such as oil palm in Ecuador, soybean in Brazil or banana in the Philippines) are given equal weight as traditional forest cover in the non-discriminatory GFC analysis (Tropek *et al.*, 2014). This critique, although valid from an ecological perspective, is unlikely to alter the mangrove data implicitly embedded in the database unless other forest types that reach the height of 5 m within the analysis period have displaced mangrove. Although displacement by forest plantations may be possible in drainage situations or at the terrestrial interface of the mangrove forest, such displacement by plantation or other forests has not been documented in the global mangrove deforestation literature which mostly attributes mangrove deforestation to displacement by aquaculture or urban expansion (Hamilton, 2013). Indeed, data integration with other sources is proposed as a method of overcoming the critique noted above (Hansen *et al.*, 2014) and this is the approach taken in this paper.

### *Mangrove Forests of the World*

MFW processes over 1000 Landsat scenes using a hybrid unsupervised and supervised classification approach (Giri *et al.*, 2011). It does not attempt to depict forest change over time but is does provide a one-time global snapshot of mangrove forest cover in the year 2000. As opposed to the continuous tree cover approach, MFW of the world provides mangrove presence or absence data at the minimum mapping unit of 1 ha. This dataset provides the second database to help delineate mangrove forest cover in this paper.

### *Terrestrial Ecoregions of the World*

TEOW is an integrated map product developed over 10-years that delineates 825 global ecoregions, nesting them within 14 biomes and 8 biogeographic realms (Olson *et al.*, 2001). "Ecoregions are relatively large units of land containing distinct assemblages of natural communities and species, with boundaries that approximate the original extent of natural communities prior to major land-use change" (Olson *et al.*, 2001 p. 933). The ecoregion framework presented has become one of the foundational geospatial layers used in biodiversity and conservation. As opposed to other land cover / land use designations, this dataset explicitly delineate the mangrove ecosystem as a unique biome in their dataset. Although, the mangrove biome does not necessarily mean mangrove is present, combined with other datasets such the depiction of whole-system mangrove biomes forest transition can be analyzed for transitions in canopy cover. This dataset provides the third database to delineate mangrove forest area, including mangrove loss and gain.

Data validation reports, measures of potential error and a comparison between continuous measures of mangrove cover vs. binary measures of cover are provided in the supplemental methods.

**RESULTS**

Mangroves are located in 105 countries (Supplemental Table 1, as well as in the special administrative areas of China (Hong Kong and Macau), the four French overseas provinces of Martinique, Guiana, Guadeloupe and Mayotte as well as the contested area of Somaliland. For reporting purposes Hong Kong and Macau are aggregated into China, the French provinces are aggregated into France and Somaliland is aggregated into Somalia. Omitted forests constitute less than 0.01% of the global mangrove total and are discussed in detail in the supplemental methods. The top 20 mangrove holding nations contain between 80% and 85% of global mangrove stocks and are presented in Table 2.

**Mangrove forests of the world (MFW) results**

Our new estimate of mangrove area, within the area identified by MFW, revised for percentage cover as opposed to presence or absence, for the year 2000 is 83,495 km$^2$ (Supp. Table 1). This is a decrease of 54,360 km$^2$ from the 137,760 km$^2$ total reported by Giri *et al.* (2011). This decrease of 39% from MFW is primarily due to a differing definition of mangrove used in the two analyses and does not evidence a substantial loss of mangrove or any error by either set of authors. Such a substantial difference in area between the two methods does suggest that binary pixel measures may indeed be inadequate for many mangrove analyses such as establishing mangrove carbon stocks for REDD programs. The difference between CGMFC-21 and nationally reported statistics compiled by the FAO (Table 1) is closer to a 50% reduction in mangrove forest cover. This is consistent with wider forest findings outside of the mangrove biome in Latin America that report 50% lower areas when using continuous remote sensing data as opposed to national estimates without remotely sensed data (DeFries *et al.*, 2002).

Mangrove forests that existed in 2000 have decreased by 1646 km$^2$ globally between 2000 and 2012 (Fig 1). This corresponds to a total loss over the analysis period of 1.97% from the year 2000 baseline. This equates to a loss globally during this period of 137 km$^2$ or 0.16% annually. The losses appear generally consistent across the period analyzed with an almost linear relationship (r$^2$ = .99) between year and loss.[1] This consistent trend with little deviation allows future trends to be reliably extrapolated from the dataset with a high amount of certainty. Extrapolated to 2014, global mangroves are estimated cover 81,484 km$^2$ (Supp. Table 1).

Myanmar appears to represent the current hotspot of mangrove deforestation with a rate of deforestation more than four times higher than the global average (Supp. Table 1). Although Myanmar has the highest rate of loss, Indonesia has by far the largest area loss. The 3.11% mangrove loss in Indonesia equates to 749 km$^2$ of mangrove loss and constitutes almost half of all global mangrove deforestation. The majority of this loss is occurring in the provinces of Kalimantan Timur and Kalimantan Selatan with a distinct deforestation hotspot visible along the eastern coast of Kalimantan. Southeast Asia has experienced relatively high amounts of loss and this is of importance as this these nations contain almost half of the global mangrove area. Other countries outside of Southeast Asia that have sustained significant mangrove losses as a percentage of their 2000 total include India and Guatemala. Within the Americas, Africa and Australia the deforestation of mangrove is approaching zero with nominal rates in many countries.

---

[1] $y = -142.27x + 83615$, where x is the last two digits of the year +1.

**Mangrove biome (TEOW) results**

Mangrove loss patterns in the entire mangrove biome exhibit mostly similar patterns to the MFW loss patterns described above, but with some important differences. Mangrove biome treecover declined from 173,067 km$^2$ in 2000 to 167,387 km$^2$ in 2012 (Supp. Table 1). We extrapolate these numbers to estimate treecover of 163,925 km$^2$ in 2014. The global deforestation rate in the mangrove biome from 2000 to 2012 is 4.73% with an annual rate of loss of 0.39% (Supp. Table 1). This indicates that the wider mangrove biome may be under more stress than the actual trees delineated as mangrove in year 2000 by MFW. Myanmar, Indonesia, Malaysia, Cambodia and Guatemala (Supp. Table 1) all have relatively high levels of tree loss within the mangrove biome. Again, Southeast Asia is the region of most concern averaging 8.08% mangrove loss during the analysis period. Significant mangrove holding nations such as Nigeria, Venezuela, Bangladesh and Fiji have established stable forest cover in the mangrove biome with loss rates close to zero during the analysis period.

**Ramsar and protected sites**

Ramsar sites and protected areas are included in the results to demonstrate the capability of our dataset to provide sub-national estuarine specific data from 2000 to present as well as provide important insights into the role of protected areas in conserving mangrove forests. Table 2 represents the almost negligible loss in the selected Ramsar areas, aside from the Everglades. The percentage of mangrove loss within the selected Ramsar sites is 50% lower than the global mangrove loss average (Table 3), with a mangrove loss rate of 0.08% annually between 2000 and 2012. The percentage of mangrove loss within all global protected areas as defined by the World Database on Protected Areas (IUCN & UNEP, 2013), using the TEOW method, is again almost 50% lower than the global average with losses of 0.21% annually between 2000 and 2012.

Mangrove area GIS raster results data (MFW) for 2000 to 2012 can be downloaded from http://dx.doi.org/10.7910/DVN/HKGBGS. Mangrove area GIS raster results data (Biome) for 2000 to 2012 can be downloaded from http://dx.doi.org/10.13016/M2ZT44. Extended tabular information can be downloaded from http://dx.doi.org/10.7910/DVN/HS5OXF.

**DISCUSSION**

This paper has presented a systematic data synthesis approach to providing continuous measures of mangrove cover utilizing the highest spatiotemporal resolutions available. The methodology designed can be applied to other forest types globally, enabling relatively rapid forest change metrics at high spatiotemporal resolutions. Utilizing continuous data has reduced the mangrove area by approximately 40% from earlier estimates. This is not a cause for concern as the difference is due to an enhanced measure of mangrove cover as opposed to a substantial loss in mangrove forest. Indeed, if we convert these data back to presence or absence the mangrove area is in very close agreement with other mangrove datasets at the country scale. The continuous mangrove variable used in this paper should provide an improved measure of mangrove when the concern is woody biomass, carbon storage and habitat degradation.

The post-2000 mangrove deforestation trend of between 0.16% and 0.39% annually represents a significant decrease in annual mangrove loss rates when compared to the proceeding decades. For example, using a synthesis of FAO data the best estimates for annual losses during the 1980s is 0.99% annually (FAO, 2007), and for the 1990s is 0.70% annually (FAO, 2007). While still suffering a substantial decline, the reported decrease in the mangrove deforestation since 2000 ameliorates the potential of a world without functional mangroves within 100-years idea that gained traction in mid-2000s (Duke *et al.*, 2007). Such concerns were based on extrapolated data from estimates of mangrove deforestation obtained from the 1980s and 1990s and the trends in these datasets appear not to have continued into the 21st Century.

The data presented address the well-documented problems of establishing consistent PES baselines and provide much needed degradation information as well as deforestation information. Mangrove carbon stock estimates, as well as the economic value placed on such carbon holdings, are enhanced by providing systematic measures of mangrove holdings at annual intervals as opposed to utilizing latitudinal estimates of carbon from singular snapshots of mangrove cover utilizing presence or absence data. These data provide systematic global estimates of mangrove cover as well as providing both the temporal and spatial resolution required for high fidelity analyses of mangrove change. Additionally, the methodology provided allows researchers to develop PES baseline and degradation products at high spatiotemporal resolutions for other forest types globally.

Although global mangrove losses have slowed considerably, and can be considered static in many nations including internationally important internationally important Ramsar sites and protected areas, this condition is not universal and Southeast Asia remains a region of concern and the discovery of Myanmar as a mangrove deforestation frontier since 2000 requires further research. Aquaculture has expanded substantially in Myanmar since 1999 (Hishamunda *et al.*, 2009) and this may be the driving force behind the deforestation although rice cultivation is additionally noted as a major driver of mangrove loss in the Ayerwaddy Delta region of Myanmar (Webb *et al.*, 2014). Indonesia remains a country of concern with annual mangrove deforestation approximately double the global average and this equates to almost half of all global mangrove losses (Supp. Table 1). These data do not elucidate on the cause of deforestation and a regional analysis is required to fully account for these losses.

In summary, the global pattern of mangrove deforestation during since 2000 is one of decreasing rates of deforestation; many nations are essentially stable, with the exception of the largest mangrove holding nations of Southeast Asia. Although the global, national and regional mangrove holdings reported in this paper are significant to the wider research community, including those interested in climate change it is the presentation of a global, systematic, continuous, annual, high resolution mangrove dataset that this research has the most utility. Researchers studying such important mangrove related issues as fisheries, conservation, $CO_2$ emissions, carbon sequestration and livelihoods now have access to the data required to undertake robust analyses into these important mangrove research questions.

## ACKNOWLEDGEMENTS

Thank you to the originators of all three input datasets as well as the supporters of the remote sensing systems that are used to build such datasets. Thank you to the students of INTR 204 at William and Mary for testing these methods. Thank you to Dan Freiss of the mangrove lab in Singapore and the Moore Foundation for reviewing drafts of this paper. Thank you to the reviewers and editors at GEB for help editing the paper.

## REFERENCES

Aizpuru, M., Achard, F. & Blasco, F. (2000) Global Assessment of Cover Change of the Mangrove Forests Using Satellite Imagery at Medium to High Resolution. In: *EEC Research Project No. 15017-1999-05 FIED ISP FR*. Joint Research Center, Ispra, Italy.

Angelsen, A. (2008) REDD models and baselines. *International Forestry Review*, **10**, 465-475.

Angelsen, A., Brockhaus, M., Sunderlin, W.D. & Verchot, L.V. (2012) *Analysing REDD+: Challenges and Choices*. Center for International Forestry Research (CIFOR), Bogor, Indonesia.

Asner, G.P., Knapp, D.E., Broadbent, E.N., Oliveira, P.J., Keller, M. & Silva, J.N. (2005) Selective logging in the Brazilian Amazon. *Science*, **310**, 480-482.

Asner, G.P., Powell, G.V., Mascaro, J., Knapp, D.E., Clark, J.K., Jacobson, J., Kennedy-Bowdoin, T., Balaji, A., Paez-Acosta, G. & Victoria, E. (2010) High-resolution forest carbon stocks and emissions in the Amazon. *Proceedings of the National Academy of Sciences*, **107**, 16738-16742.

Barbier, E.B. (2006) Mangrove Dependency and the Livelihoods of Coastal Communities in Thailand. *Environment and Livelihoods in Tropical Coastal Zones: Managing Agriculture-Fishery-Aquaculture Conflicts (Comprehensive Assessment of Water Management in Agriculture Series)* (ed. by C.T. Hoanh, T.P. Tuong, J.W. Gowing and B. Hardy). Oxford University Press, London.

Barbier, E.B. & Cox, M. (2004) An Economic Analysis of Shrimp Farm Expansion and Mangrove Conversion in Thailand. *Land Economics*, **80**, 389-407.

Bennett, B. (2001) What is a Forest? On the Vagueness of Certain Geographic Concepts. *Topoi*, **20**, 189-201.

Bouillon, S., Borges, A.V., Castañeda-Moya, E., Diele, K., Dittmar, T., Duke, N.C., Kristensen, E., Lee, S.Y., Marchand, C., Middelburg, J.J., Rivera-Monroy, V.H., Smith, T.J. & Twilley, R.R. (2008) Mangrove Production and Carbon Sinks: A Revision of Global Budget Estimates. *Global Biogeochemical Cycles*, **22**, GB2013.

Chong, V.C. (2007) Mangroves-Fisheries Linkages in the Malaysian Perspective. *Bulletin of Marine Science*, **80**, 755-772.

DeFries, R.S., Houghton, R.A., Hansen, M.C., Field, C.B., Skole, D. & Townshend, J. (2002) Carbon Emissions from Tropical Deforestation and Regrowth Based on Satellite Observations for the 1980s and 1990s. *Proceedings of the National Academy of Sciences of the United States of America*, **99**, 14256-14261.

DeFries, R.S., Field, C.B., Fung, I., Justice, C.O., Los, S., Matson, P.A., Matthews, E., Mooney, H.A., Potter, C.S. & Prentice, K. (1995) Mapping the land surface for global atmosphere-biosphere models: Toward continuous distributions of vegetation's functional properties. *Journal of Geophysical Research: Atmospheres (1984–2012)*, **100**, 20867-20882.

Donato, D.C., Kauffman, J.B., Murdiyarso, D., Kurnianto, S., Stidham, M. & Kanninen, M. (2011) Mangroves among the Most Carbon-Rich Forests in the Tropics. *Nature Geoscience*, **4**, 293-297.

Duke, N.C., Meynecke, J.O., Dittmann, S., Ellison, A.M., Anger, K., Berger, U., Cannicci, S., Diele, K., Ewel, K.C., Field, C.D., Koedam, N., Lee, S.Y., Marchand, C., Nordhaus, I. & Dahdouh-Guebas, F. (2007) A World without Mangroves? *ScienceMag*, **317**, 41-42.

FAO (1995) Forest Resources Assessment 1990. In: *Global Forest Resources Assessment*. FAO, Rome, Italy.

FAO (1997) *State of the World's Forests, 1997*. Words and Publications, Oxford, UK.

FAO (2000) Global Forest Resources Assessment. In. FAO, Rome, Italy.

FAO (2003) Status and trends in mangrove area extent worldwide. In: *Forest Resources Assessment Programme. Working Paper 63* eds. M.L. Wilkie and S. Fortuna). FAO, Rome, Italy.

FAO (2005) Global Forest Resources Assessment 2005. In: *Global Forest Resources Assessment*. FAO, Rome, Italy.

FAO (2007) The World's Mangroves 1980-2005. In: *FAO Forestry Paper*. FAO, Rome.

FAO (2010) Global Forest Resources Assessment 2010. In: *Global Forest Resources Assessment*. FAO, Rome, Italy.

FAO Fisheries and Aquaculture Department (2004) Mangrove Forest Management Guidelines. In: *FAO Forestry Paper 117* (ed. F.F.a.A. Department), p. 319. FAO Fisheries and Aquaculture Department, Rome.

Fisher, P. & Spalding, M. (1993) Protected Areas with Mangrove Habitat. In, p. 60. World Conservation Centre, Cambridge, UK.

Friess, D. & Webb, E. (2011) Bad data equals bad policy: how to trust estimates of ecosystem loss when there is so much uncertainty? *Environmental Conservation*, **38**, 1-5.

Friess, D.A. & Webb, E.L. (2014) Variability in mangrove change estimates and implications for the assessment of ecosystem service provision. *Global Ecology and Biogeography*,

Giri, C., Ochieng, E., Tieszen, L.L., Zhu, Z., Singh, A., Loveland, T., Masek, J. & Duke, N. (2011) Status and Distribution of Mangrove Forests of the World Using Earth Observation Satellite Data. *Global Ecology and Biogeography*, **20**, 154-159.

Griscom, B., Shoch, D., Stanley, B., Cortez, R. & Virgilio, N. (2009a) Sensitivity of amounts and distribution of tropical forest carbon credits depending on baseline rules. *Environmental Science & Policy*, **12**, 897-911.

Griscom, B., Shoch, D., Stanley, B., Cortez, R. & Virgilio, N. (2009b) Implications of REDD baseline methods for different country circumstances during an initial performance period. *The Nature Conservancy, Arlington*, 1-35.

Groombridge, B. (1992) Global Biodiversity: Status of the Earth's Living Resources. In: eds. Unep, Wcmc, T.N.H. Museum, Iucn, W.F.F. Nature and W.R. Institute), London.

Hamilton, S. (2013) Assessing the Role of Commercial Aquaculture in Displacing Mangrove Forest. *Bulletin of Marine Science*, **89**, 585-601.

Hansen, M.C., Potapov, P.V., Margono, B., Stehman, S., Turubanova, S.A. & Tyukavina, A. (2014) Response to Comment on High-resolution global maps of 21st-century forest cover change. *Science*, **344**

Hansen, M.C., Potapov, P.V., Moore, R., Hancher, M., Turubanova, S.A., Tyukavina, A., Thau, D., Stehman, S.V., Goetz, S.J., Loveland, T.R., Kommareddy, A., Egorov, A., Chini, L., Justice, C.O. & Townshend, J.R.G. (2013) High-Resolution Global Maps of 21st-Century Forest Cover Change. *Science*, **342**, 850-853.

Hishamunda, N., Bueno, P.B., Ridler, N. & Yap, W.G. (2009) *Analysis of aquaculture development in Southeast Asia*. FAO.

Homer, C.H., Fry, J.A. & Barnes, C.A. (2012) The national land cover database. *US Geological Survey Fact Sheet*, **3020**, 1-4.

Huettner, M., Leemans, R., Kok, K. & Ebeling, J. (2009) A comparison of baseline methodologies for 'Reducing Emissions from Deforestation and Degradation'. *Carbon Balance and Management*, **4**, 4.

ITTO & ISME (1993) Mangrove Ecosystems: Technical Reports. *ITTO/ISME/JIAM Project PD71/8* (ed by International Society for Mangrove Ecosystems (Isme), International Tropical Timber Organization (Itto) and J.I.a.F.M. (Jiam)). Nishihara, Japan.

IUCN & UNEP (2013) The World Database on Protected Areas - WDPA. In. UNEP, Cambridge, UK.

Lanly, J.P. (1982) Tropical forest resources assesment. In: *Global Forest Resources Assessment* (ed. Fao). FAO, Rome, Italy.
20

Lugendo, B.R., Nagelkerken, I., Kruitwagen, G., van der Velde, G. & Mgaya, Y.D. (2007) Relative Importance of Mangroves as Feeding Habitats for Fishes: A Comparison Between Mangrove Habitats with Different Settings. *Bulletin of Marine Science*, **80**, 497-512.

Murdiyarso, D., Purbopuspito, J., Kauffman, J.B., Warren, M.W., Sasmito, S.D., Donato, D.C., Manuri, S., Krisnawati, H., Taberima, S. & Kurnianto, S. (2015) The potential of Indonesian mangrove forests for global climate change mitigation. *Nature Climate Change*, **5**, 1089-1092.

Olson, D.M., Dinerstein, E., Wikramanayake, E.D., Burgess, N.D., Powell, G.V.N., Underwood, E.C., D'amico, J.A., Itoua, I., Strand, H.E., Morrison, J.C., Loucks, C.J., Allnutt, T.F., Ricketts, T.H., Kura, Y., Lamoreux, J.F., Wettengel, W.W., Hedao, P. & Kassem, K.R. (2001) Terrestrial Ecoregions of the World: A New Map of Life on Earth: A new global map of terrestrial ecoregions provides an innovative tool for conserving biodiversity. *BioScience*, **51**, 933-938.

Ruiz-Luna, A., Acosta-Velázquez, J. & Berlanga-Robles, C.A. (2008) On the reliability of the data of the extent of mangroves: A case study in Mexico. *Ocean & Coastal Management*, **51**, 342-351.

Saenger, P., Hegerl, E.J. & Davie, J.D.S. (1983) Global Status of Mangrove Ecosystems. In: *Commission on Ecology Papers No. 3*, p. 88. World Conservation Union (IUCN), Gland, Switzerland.

Satapathy, D.R., Krupadam, R.J., Kumar, L.P. & Wate, S.R. (2007) The Application of Satellite Data for the Quantification of Mangrove Loss and Coastal Management in the Godavari Estuary, East Coast of India. *Environmental Monitoring and Assessment*, **134**, 453-469.

Siikamäki, J., Sanchirico, J.N. & Jardine, S.L. (2012) Global economic potential for reducing carbon dioxide emissions from mangrove loss. *Proceedings of the National Academy of Sciences*, **109**, 14369-14374.

Spalding, M., Blasco, F. & Field, C. (1997) *World Mangrove Atlas*. International Society for Mangrove Ecosystems, Okinawa, Japan.

Spalding, M., Kainuma, M. & Collins, L. (2010) *World Atlas of Mangroves*. Earthscan, London, UK.

Tropek, R., Sedláček, O., Beck, J., Keil, P., Musilová, Z., Šímová, I. & Storch, D. (2014) Comment on "High-resolution global maps of 21st-century forest cover change". *Science*, **344**, 981-981.

Webb, E.L., Jachowski, N.R.A., Phelps, J., Friess, D.A., Than, M.M. & Ziegler, A.D. (2014) Deforestation in the Ayeyarwady Delta and the conservation implications of an internationally-engaged Myanmar. *Global Environmental Change*, **24**, 321-333.

**TABLES**

Table 1. Global mangrove area estimates in km$^2$ by year and author. The mangrove area estimates within each decade are highly variable.

| ID | Source / Citation / Page | Reference Year | Countries | Mangrove Area (km$^2$) |
|---|---|---|---|---|
| 1 | FAO (FAO, 2007), p. 9. | 1980 | Global | 187,940 |
| 2 | Lanjly (Lanly, 1982), p. 43. | 1980 | 76 | 154,620 |
| 3 | Saenger (Saenger *et al.*, 1983), p. 11-12. | 1983 | 66 | 162,210 |
| 4 | FAO (FAO, 2004), Table 2.3. | 1980-1985 | 56 | 165,300 |
| | **1980s Mean (1-4)** | | | **167,518** |
| 5 | FAO (FAO, 2007), p. 9. | 1990 | Global | 169,250 |
| 6 | Groombridge (Groombridge, 1992), p. 325-326. | 1992 | 87 | 198,478 |
| 7 | ITTO / ISME (ITTO & ISME, 1993), p. 6. | 1993 | Global | 141,973 |
| 8 | Fisher (Fisher & Spalding, 1993), p. 11. | 1993 | 91 | 198,817 |
| 9 | Spalding (Spalding *et al.*, 1997), p. 23. | 1997 | 112 | 181,077 |
| | **1990s Mean (5-9)** | | | **177,919** |
| 10 | Spalding (Spalding *et al.*, 2010), p. 6. | 2000-2001 | 123 | 152,361 |
| 11 | FAO (FAO, 2007), p. 9. | 2000 | Global | 157,400 |
| 12 | Aizpuru (Aizpuru *et al.*, 2000), secondary source. | 2000 | 112 | 170,756 |
| 13 | Giri (Giri *et al.*, 2011), p. 156. | 2000 | Global | 137,600 |
| 14 | FAO (FAO, 2007), p. 9. | 2005 | Global | 152,310 |
| | **2000s Mean (10-14)** | | | **154,085** |

Table 2. The top 20 mangrove holding nations as of 2000 and their change in mangrove area from 2000 to 2014 in km² and percentage of global total. The top 4 countries contain greater than 49% of the world's mangroves.

| 2000 MFW Rank | Country Name | 2000 MFW km² | 2000 MFW Percent | 2000 BIOME km2 | 2000 BIOME Percent | 2014 MFW km2 | 2014 MFW Percent | 2014 BIOME km2 | 2014 BIOME Percent |
|---|---|---|---|---|---|---|---|---|---|
| 1 | Indonesia | 24073 | 28.83% | 46642 | 26.95% | 23143 | 28.40% | 42278 | 25.79% |
| 2 | Brazil | 7721 | 9.25% | 18168 | 10.50% | 7663 | 9.40% | 17287 | 10.55% |
| 3 | Malaysia | 4969 | 5.95% | 8738 | 5.05% | 4691 | 5.76% | 7616 | 4.65% |
| 4 | Papua New Guinea | 4190 | 5.02% | 5982 | 3.46% | 4169 | 5.12% | 6236 | 3.80% |
| 5 | Australia | 3327 | 3.98% | 3359 | 1.94% | 3315 | 4.07% | 3314 | 2.02% |
| 6 | Mexico | 3021 | 3.62% | 6240 | 3.61% | 2985 | 3.66% | 6036 | 3.68% |
| 7 | Myanmar | 2793 | 3.34% | 4205 | 2.43% | 2508 | 3.08% | 3783 | 2.31% |
| 8 | Nigeria | 2657 | 3.18% | 6944 | 4.01% | 2653 | 3.26% | 6908 | 4.21% |
| 9 | Venezuela | 2416 | 2.89% | 7579 | 4.38% | 2401 | 2.95% | 7516 | 4.59% |
| 10 | Philippines | 2091 | 2.50% | 2115 | 1.22% | 2060 | 2.53% | 2084 | 1.27% |
| 11 | Thailand | 1933 | 2.32% | 4362 | 2.52% | 1876 | 2.30% | 3936 | 2.40% |
| 12 | Bangladesh | 1774 | 2.12% | 2317 | 1.34% | 1773 | 2.18% | 2314 | 1.41% |
| 13 | Colombia | 1674 | 2.01% | 6313 | 3.65% | 1672 | 2.05% | 6236 | 3.80% |
| 14 | Cuba | 1660 | 1.99% | 2471 | 1.43% | 1624 | 1.99% | 2407 | 1.47% |
| 15 | United States | 1612 | 1.93% | 1616 | 0.93% | 1553 | 1.91% | 1554 | 0.95% |
| 16 | Panama | 1328 | 1.59% | 2768 | 1.60% | 1323 | 1.62% | 2673 | 1.63% |
| 17 | Mozambique | 1226 | 1.47% | 2716 | 1.57% | 1223 | 1.50% | 2658 | 1.62% |
| 18 | Cameroon | 1119 | 1.34% | 1344 | 0.78% | 1113 | 1.37% | 1323 | 0.81% |
| 19 | Gabon | 1087 | 1.30% | 3929 | 2.27% | 1081 | 1.33% | 3864 | 2.36% |
| 20 | Ecuador | 938 | 1.12% | 1971 | 1.14% | 935 | 1.15% | 1906 | 1.16% |
|  | TOTAL | 71608 | 85.76% | 139777 | 80.76% | 69761 | 85.61% | 131931 | 80.48% |

Table 3. Mangrove loss by Ramsar site.

Mangrove loss from 2000 to 2012 in km2 and percent of 2000 mangrove area, for specific Ramsar wetlands on each continent in areas with mangrove present in 2000.

| Site | 2000 | 2012 | Percent Loss |
|---|---|---|---|
| **Sundarbans** | 197,994 | 197,961 | 0.02% |
| **Everglades** | 93,090 | 89,945 | 3.38% |
| **Douala Edéa** | 24,648 | 24,532 | 0.47% |
| **Cayapas-Mataje** | 14,807 | 14,748 | 0.40% |
| **Garig Gunak Barlu** | 11,360 | 11,296 | 0.56% |
| TOTAL | 341,899 | 338,482 | 1.00% |

**Supplemental Table 1**

Please see the http://dx.doi.org/10.7910/DVN/HS5OXF for this table.

*This table represents the full result set for all countries within mangrove measures for 2000 to 2012 with mangrove estimates for 2013 and 2014. All 105 nations are included with annual measures of mangrove cover for each nation and each nation's percentage of the global mangrove total. Additionally the table represents entire change percentages as well as annual change percentages*

**FIGURES**

Figure 1.

Global mangrove change from 2000 to 2012 with extrapolated data for 2014. Grey represents all locations with known mangrove existing in 2000 and black represents all locations in the wider mangrove biome. The x-axis is year and the y-axis is km$^2$.

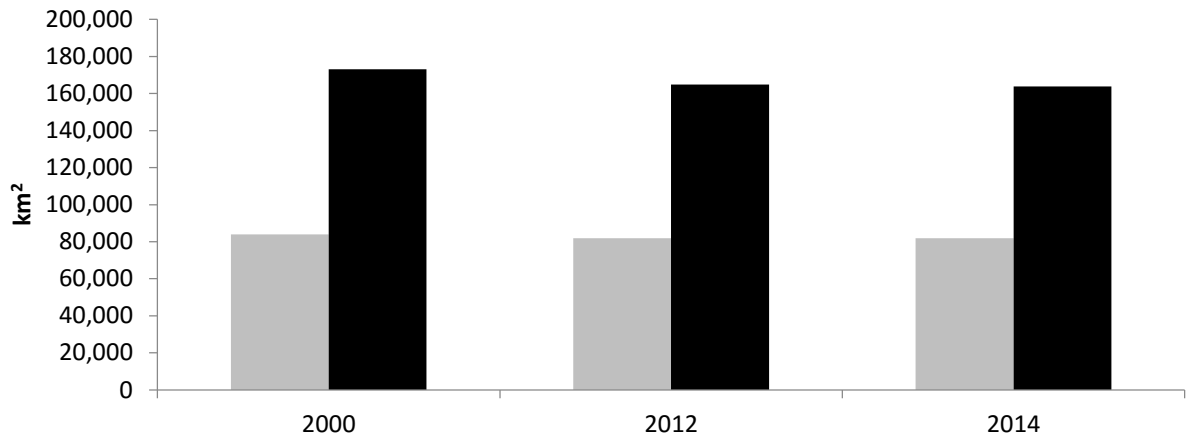

# SUPPLEMENTAL METHODS

## Global Forest Cover

The base year for GFC is 2000 and the data can be accessed, visualized, and downloaded in raster GIS format from Google Earth Engine. The GFC global accuracy level is reported as 99.6% (n = 1500, 0.7) for areas of forest loss or no loss, and 99.7% (n = 1500, 0.6) for areas of forest gain or no gain. Within the tropical regions these values change to 99.5% ( n = 628, 0.1) and 99.7% (n = 628, 0.1) respectively (Hansen *et al.*, 2013). GFC was provided in raster format at the native x, y resolution of 2.7777 × 10-4° in the non-projected WGS 1984 coordinate system EPSG: 4326.

Manipulating big-data such as GFC is challenging and one million CPU hours were required and Google provided a parallelism solution to overcome the processing issues (Hansen *et al.*, 2013). Although not as substantial as the original data creation, the manipulation of the dataset for mangrove forests required deconstructing the data into multiple tiles, processing the data on octadic-cores, and reconstituting the data when each tile was complete. To achieve this end we subdivided the Hansen data into 10° by 10° tiles covering the entire possible longitudinal mangrove range and the mangrove forested latitudes between 40°N and 30°S. Once tiles with no land or no recorded evidence of mangrove were removed this resulted in 105 tiles of data consisting of 1.36 billion potential locations of mangrove tree-cover across the entire dataset. Each location's tree measure can be 1% - 100% with 1 indicating only 1% of the pixel has tree cover and 100 indicating the entire pixel is forested. A value of zero indicates deforestation of mangrove from an earlier period. This increases to 137.51 billion the potential mangrove tree measures globally.

## Mangrove Forests of the World

MFW was provided as a large multipart vector polygonal feature class with the polygon outlines representing the outline of the input landsat pixels (where mangrove is present) in the non-projected WGS 1984 coordinate system EPSG: 4326. The MFW dataset can be visualized in Google Earth Engine and the data can be downloaded in vector GIS format from UNEP-WCMS Ocean Data Viewer. 137,760 km$^2$ of mangrove existed and this was less than other estimates of this time. MFW has a published RMSE of ± 1/2 pixel. Despite WMF clearly having some alignment and omission errors, such errors appear nominal and will have a negligible effect on global, national or even localized estuarine specific mangrove estimates.

## Terrestrial Ecoregions of the World

TEOW does not provide error metrics and the data were provided in vector feature format with a vector polygon representing each portion of the mangrove biome in the non-projected WGS 1984 coordinate system EPSG: 4326. The TEOW dataset can be downloaded in GIS format from the WWF.

## Data Validation

Supplemental Figure 1. Histogram of Florida Mangrove Difference (CGMFC-21 MF 2011 vs. NLCD 2011)

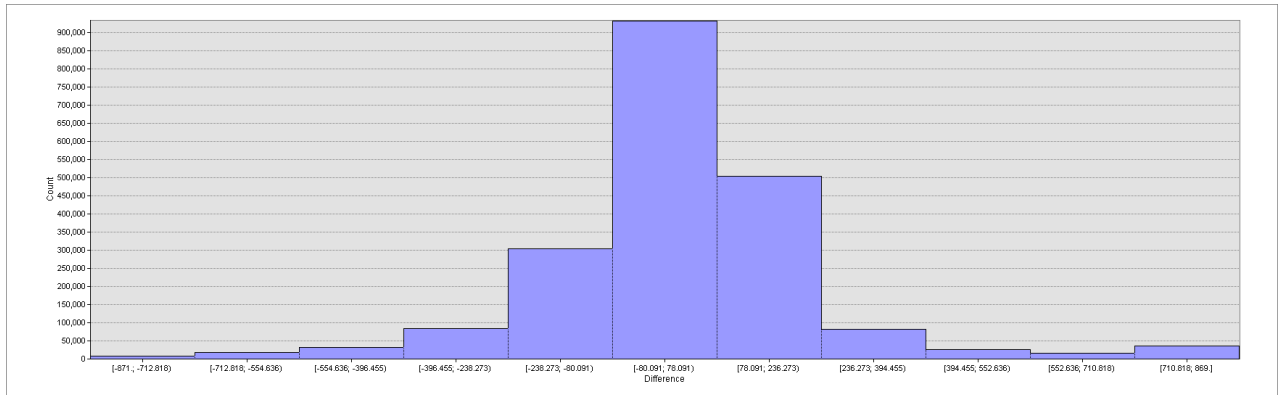

*The pixel difference between the two datasets clusters around the mean difference of 35m and shows a normal distribution.*

## Continuous Measures of Mangrove Forest.

Figure 2 demonstrates the difference between traditional binary mangrove calculations and the continuous mangrove cover utilized in this analysis. The region depicted is a degraded area of mangrove forest in West Africa. The lower panel depicts mangrove presence only and the area of forest is calculated to be 487,869,714 m2. The upper panel is the same aerial extent and the same pixel coverage but with percentage of pixel forested applied. The calculated mangrove area is now reduced to 179,736,766 m2 in the continuous representation of mangrove cover. The majority of pixels have less than 40% mangrove cover and this substantially alters the mangrove forest area calculation. Other potential causes of difference are the omission of scrub or juvenile mangrove from CGMFC-21. Such omissions may be particularly important towards the fringes of mangrove cover at the sub-tropics in such nations as New Zeeland and the United States, although the Florida analysis presented earlier and digital mangrove maps obtained from Land Information New Zealand both indicate that the mangrove measures produced by CGMFC-21 are highly representative of the mangrove information contained within other national databases at the latitudinal edges of global mangrove presence.

Supplemental Figure 2. A comparison of year 2000 presence and absence mangrove data as opposed to continuous mangrove cover.

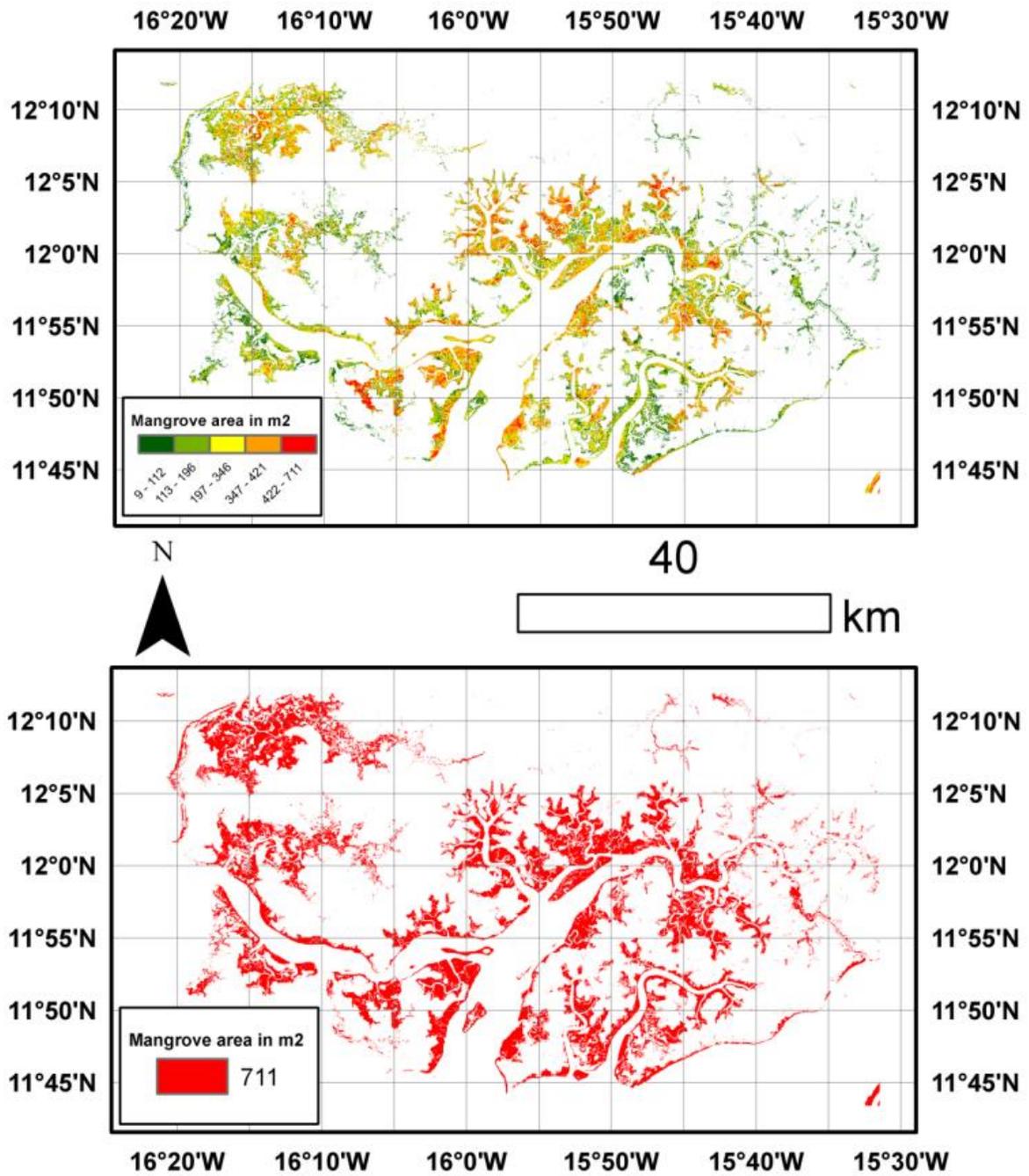

*The upper panel represents a region from CGMFC-21 with continuous measurements and the lower panel represents the same time and place but utilizing mangrove presence or absence.*

**Omitted Sites**

American Samoa, Samoa, Guam, Northern Mariana Islands, Marshall Islands, Nauru and Tonga are excluded from this analysis as no reliable Landsat L1T data exists for compilation of GFC on these smaller islands. We estimate the mangrove total within these areas to be 52 ha (FAO, 2007), 370 ha (FAO, 2007), 70 ha (FAO, 2007), 7 ha (FAO, 2007), 4 ha (Gilman *et al.*, 2006), 2 ha (FAO, 2007), and 1305 ha (FAO, 2007) respectively for a combined total of 18.1 km$^2$ or < 0.01% of the global mangrove total (FAO, 2007). Mangroves in Egypt, Kuwait, Bahrain, Wallis and Futuna, and Tokelau are additionally excluded as evidence suggests that these mangroves are all under 5 m tall (FAO, 2005, 2007). We estimate the combined mangrove area of these countries to be approximately 0.1 km$^2$ constituting a negligible amount of the global total. The FAO (2007), additionally estimates 0.1 km$^2$ of mangrove in Dominica in 1991 and this is supported in the academic literature (Godt, 1990) and by photographic evidence but these mangroves are missing in MFW and TEOW and are omitted in this analysis. Approximately 20 km$^2$ of mangroves in Mauritius are documented in the academic literature (Appadoo, 2003) although FAO (FAO 2007) reports only 1.2 km$^2$. The majority of these mangroves are likely under 5 m (Appadoo, 2003) but again are missing in MFW and TEOW and are omitted in this analysis. A nominal 0.05 km$^2$ amount of mangrove is reported in Montserrat (FAO, 2007) but are missing in MFW and TEOW and are omitted in this analysis.

The reported approximately 3000 ha of mangrove in Niue (Ellison, 1999; FAO, 2003, 2007) are excluded from the analysis as these are no longer considered mangrove forests (Spalding *et al.*, 1997; Gilman *et al.*, 2006). Recent planting projects in Kiribati and Tuvalu may have added to the mangrove total towards the end of the study period although the planting dates suggest these mangroves will still be in the juvenile stage and under 5 m in height, additionally many of the mangrove planting schemes appear to have suffered almost 100% mortality (Baba *et al.*, 2009). An unknown quantity of mangroves in São Tomé and Príncipe are reported in various sources (FAO, 2003, 2007; Ministry for Natural Resources and the Environment, 2007) but are omitted from MFW and hence are excluded in this analysis. The combined estimate of all omitted mangroves, even when taking the highest estimate of cover in each location, is approximately 0.01% of the reported global total mangrove area (FAO, 2007).

MFW is missing data for Congo and the Angolan exclave of Cabinda in the distributed GIS file although these data are present in other delivery formats, such as Google Earth Engine. Reliable estimates of area and percentage cover require the GIS file. For Congo we utilized the 2000 FAO estimate (FAO 2007) and multiplied by 0.605 that is the global adjustment of change between whole pixel values and continuous forest cover values for 2000 obtained from results in this paper. For Cabinda we estimate from Google Earth Engine that the province contains 15% of the Angolan total of mangrove forest and adjust the data accordingly.

**Supplemental References**

Appadoo, C. (2003) Status of mangroves in Mauritius. Journal of Coastal Development, 7, 1-4.

Baba, S., Nakao, Y. & Yamagami, S. (2009) Challenges of planting mangroves in Kiribati. ISME/GLOMIS Electronic Journal, 7, 9-10.

Ellison, J. (1999) Status report on Pacific island mangroves. Marine and Coastal Biodiversity in the Tropical Island Pacific Region: Volume 2. Population, Development and Conservation Priorities (ed. by L.G. Eldredge, J.E. Maragos and P.L. Holthus), pp. 3-19. Pacific Science Association and East West Center, Honolulu.

FAO (2003) Status and trends in mangrove area extent worldwide. In: Forest Resources Assessment Programme. Working Paper 63 eds. M.L. Wilkie and S. Fortuna). FAO, Rome, Italy.

FAO (2005) Global Forest Resources Assessment 2005. In: Global Forest Resources Assessment. FAO, Rome, Italy.

FAO (2007) The World's Mangroves 1980-2005. In: FAO Forestry Paper. FAO, Rome.

Gilman, E., Van Lavieren, H., Ellison, J., Jungblut, V., Wilson, L., Areki, F., Brighouse, G., Bungitak, J., Dus, E. & Henry, M. (2006) Pacific Island mangroves in a changing climate and rising sea: UNEP Regional Seas Reports and Studies No. 179. In: UNEP Regional Seas Reports and Studies. nited Nations Environment Programme, Regional Seas Programme, Nairobi, Kenya.

Godt, M.C. (1990) Mangroves Found in Dominica (West Indies). Vegetatio, 86, 115-117.

Hansen, M.C., Potapov, P.V., Moore, R., Hancher, M., Turubanova, S.A., Tyukavina, A., Thau, D., Stehman, S.V., Goetz, S.J., Loveland, T.R., Kommareddy, A., Egorov, A., Chini, L., Justice, C.O. & Townshend, J.R.G. (2013) High-Resolution Global Maps of 21st-Century Forest Cover Change. Science, 342, 850-853.

Ministry for Natural Resources and the Environment (2007) National Report on the Status of Biodiversity in São Tomé and Príncipe In: (ed. M.F.N.R.a.T. Environment). Ministry for Natural Resources and the Environment, São Tomé and Príncipe.

Spalding, M., Blasco, F. & Field, C. (1997) World Mangrove Atlas. International Society for Mangrove Ecosystems, Okinawa, Japan.